\documentclass[aps,prb,showpacs,preprintnumbers,amsmath,amssymb,twocolumn,superscriptaddress,longbibliography]{revtex4-1}

\usepackage[dvipdfmx]{graphicx}
\usepackage{dcolumn}
\usepackage{bm}
\usepackage{color}

\usepackage{amsmath}	
\usepackage{amsfonts}
\usepackage{braket}
\usepackage{amsthm}

\usepackage{bbm}

\usepackage[export]{adjustbox}
\graphicspath{{fig/}}

\newcommand{\calP}{{\mathcal{P}}}
\newcommand{\calQ}{{\mathcal{Q}}}

\newtheorem*{claim}{Claim}

\begin{document}

\title{
Many-body multipole index and bulk-boundary correspondence
}

\author{Yasuhiro Tada}
\email[]{ytada@hiroshima-u.ac.jp}
\affiliation{
Quantum Matter Program, Graduate School of Advanced Science and Engineering, Hiroshima University,
Higashihiroshima, Hiroshima 739-8530, Japan}
\affiliation{Institute for Solid State Physics, University of Tokyo, Kashiwa 277-8581, Japan}

\author{Masaki Oshikawa}
\affiliation{Institute for Solid State Physics, University of Tokyo, Kashiwa 277-8581, Japan}

\begin{abstract}
We propose new dipole and quadrupole indices
for interacting insulators with point group symmetries.
The proposed indices are defined in terms of many-body quantum multipole operators combined with the generator
of the point group symmetry. Unlike the original multipole operators,
these combined operators commute with Hamiltonian under the symmetry 
and therefore their eigenvalues are quantized.
This enables a clear identification of nontrivial multipolar states. 
We calculate the multipole indices in representative models and show their effectiveness as
order parameters.
Furthermore, we demonstrate a bulk-boundary correspondence:
a non-zero index implies the existence of edge/corner states
under the the point group symmetry.
\end{abstract}

\maketitle

\section{introduction}

Multipoles
provide essential information on charge degrees of freedom in materials.
There are characteristic charge distributions on the surface of a material
as a result of a uniform multipole order in the bulk.
Therefore, it is naturally considered that there exists a ``bulk multipole moment" which is defined
for a system with the periodic boundary condition. 
This has been a subject of intensive research for decades,
and it is now widely recognized that 
the bulk dipole moment $P_x$ is described by Berry-Zak phase of 
a wavefunction~\cite{Resta2007,Vanderbilt2018,KingSmith1993,Vanderbilt1993,Zak1989}.
Furthermore, 
the bulk dipole moment can also be described by the dipole moment operator 
$U_x\sim e^{iP_x}$~\cite{Resta1998,Resta1999}.
The dipole moment operator $U_x$ is closely related to the Lieb-Schultz-Mattis theorem 
and it works as an order parameter of symmetry protected topological states 
in one-dimension~\cite{LSM1961,Nakamura2002}. 
Therefore, the dipole moment operator is regarded as a fundamental quantity not only for electric insulators
but also for general gapped quantum states.
Unfortunately, however, there are several subtleties in applications of the dipole operator $U_x$ to general systems.
For example, although it was proved that the argument of the expectation value $\langle U_x\rangle$ 
agrees with the dipole moment evaluated by the Berry phase formula in gapped one-dimensional systems,
such an equivalence may break down in higher dimensions~\cite{WatanabeOshikawa2018}.
This stems from the fact that $U_x$ is not a conserved charge, and thus its expectation value
can vanish in the thermodynamic limit.
While it may be still possible that the argument (phase factor) is well defined and gives the
dipole moment in the thermodynamic limit even if the expectation value vanishes,
this makes the formulation rather subtle.

Compared to the dipoles, 
bulk characterizations of higher order multipoles such as the quadrupole $Q_{xy}$ are even less understood.
There are gapless corner or hinge states in multipole insulators with open boundaries
and emergence of such gapless modes can be characterized by state-based quantities 
such as 
the nested Wilson loops, Wannier centers, and symmetry indicators 
in non- (or weakly) interacting systems
~\cite{Benalcazar2017Science,Benalcazar2017,Schindler2018,Langbehn2017,Song2017,Ezawa2018,
Ezawa2018PRL,Khalaf2018}.
For general interacting systems, ``bulk multipole operators'',
such as the bulk quadrupole moment operator $U_{xy}\sim e^{iQ_{xy}}$, 
were introduced as generalizations
of the bulk dipole operators~\cite{Kang2019,Wheeler2019}.
As in the case of the dipole operator, its ground-state expectation value generally vanishes
in the thermodynamic limit.
This leads to a subtlety in (and possibly to the ill-definedness of) the bulk operator formulation of 
the multipole moments.
Furthermore, the bulk operator formulation of the multipole moments is shown to have more
pathological features, such as the dependence on the choice of the origin~\cite{Ono2019}.
So far,
several other topological indices have been proposed for characterizations of
bulk multipole insulators~\cite{Araki2020,You2020,Kang2021,Wienand2022,Fukui2018,Zhu2020,qchem1,qchem2,Hofstadter1,Hofstadter2,shift2019},
but their relations to multipole moments are not well understood.

In this study, 
we propose new many-body indices for dipole and quadrupole insulators,  
which have natural interpretation in terms of
the response to an external electric field and are closely related to bulk multipole operators,
but are defined by exact quantum numbers of a deformed system.
As a consequence,
the indices are quantized under point group symmetries.
As another advantage, unlike the bulk multipole operators,
the new indices are compatible with the periodicity of the system,
Finally, the present formulation can describe a bulk-boundary correspondence in multipole insulators.

\section{Definition of quantized multipole index}
\begin{figure}[tbh]
\includegraphics[width=7.0cm]{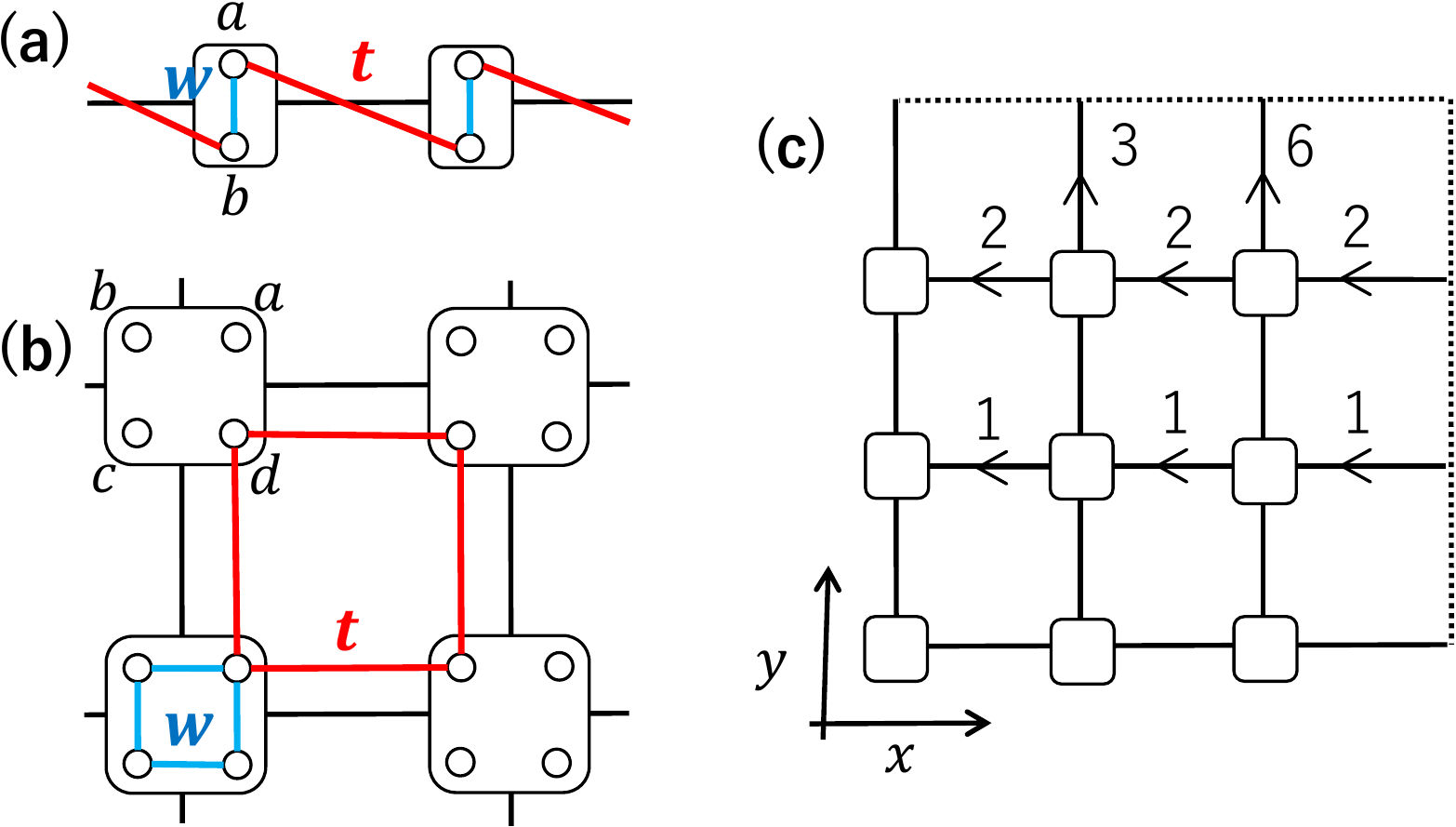}
\caption{(a) SSH model and (b) BBH model, where squares represent sites with multi-orbitals.
(c) The gauge configuration for $L=3$ with the periodic boundary condition.
Each number on the bonds corresponds to $A_{jk}$ in unit of $2\pi /L^2=2\pi/9$.
The site at the left-bottom corner is defined as the origin $(x,y)=(0,0)$.
This gauge configuration can be defined for general models~\cite{YTMO}.
}
\label{fig:model}
\end{figure}

Here we sketch the key ideas, and define novel multipole indices.
While our discussions apply to more general systems,
to be concrete, we consider the one-dimensional Su-Schrieffer-Heeger (SSH) model for dipoles~\cite{SSH1979}
and the two-dimensional Benalcazar-Bernevig-Hughes (BBH) model 
for quadrupoles~\cite{Benalcazar2017Science,Benalcazar2017}
(Fig.~\ref{fig:model} (a), (b)).
Both models can be represented by the Hamiltonian of the form 
\begin{align}
H(A)&=\sum_{jk,\mu\nu}
e^{iA_{jk}}t_{jk}^{\mu\nu}c^{\dagger}_{j\mu}c_{k\nu}+\sum_{j,\mu\nu}w_{j}^{\mu\nu}c^{\dagger}_{j\mu}c_{j\nu}+H_{\rm int}
\label{eq:HA},
\end{align}
where $t^{\mu\nu}_{jk}$ is the inter-site hopping and $w^{\mu\nu}_j$ is the intra-site hybridization between
local orbitals.
Details of $t^{\mu\nu}_{jk},w^{\mu\nu}_j$ are explained in Appendix~\ref{app:model}.
The vector potential $A_{jk}$ is an external probe field which is distinguished from phases in 
$t^{\mu\nu}_{jk},w^{\mu\nu}_j$.
We have also added the interaction 
\begin{align}
H_{\rm int}=\sum V_{jk}n_jn_k .
\end{align}
Here the particle number operator $n_j$ at site $j$ is defined as
\begin{align}
    n_j=\sum_{\mu}c^\dagger_{j\mu} c_{j\mu}-\rho,
    \label{eq:def_n}
\end{align}
where $\rho$ is the average particle number per site.
Note that we define $n_j$ differently from the standard one by subtracting the
average particle number $\rho$, for later convenience.
In this study, we focus on integer filling $\rho\in{\mathbb Z}$.
Thus $n_j \in \mathbb{Z}$ still holds.
The interaction $V_{jk}$ does not necessarily have translation symmetry, but is assumed to be point group symmetric.

We consider a finite system of linear size $L$, and impose the periodic boundary conditions.
Let $x_j, y_j$ as the $x$- and $y$-components of the coordinate of the site $j$.
The multipole moments are probed by electric fields, as follows.
In terms of the electron number operator $n_j$,
the dipole moment is given as $\calP_x = \sum_j x_j n_j$, and
the $xy$-component of the quadrupole moment is given as 
$\calQ_{xy} = \sum_j x_j y_j n_j$.
The dipole moment couples to the \emph{external} electric field through the dipole energy
\begin{align}
 H_{\text{dipole}} = - E_x \calP_x.
 \label{eq:dipole}
\end{align}
Likewise, the quadrupole moment couples to the gradient of the electric field as 
$  - \sum_{\alpha,\beta} \partial_\alpha E_\beta \calQ_{\alpha \beta}$.

Now, we can express the electric field in terms of a time-dependent vector potential $\vec{A}$, instead
of the gradient of the scalar potential~\cite{Kang2019,Wheeler2019}.
For example, in the SSH model,
we can take the ground state at $\vec{A} =0$ as the initial state, and consider the
time-dependent uniform vector potential
$A_x = ( - E_xt ) $ for $0<t<T$,
which can be interpreted as an insertion of an Aharonov-Bohm (AB) flux.
Here we set $E_x T = - 2\pi/L$, where $L$ is the linear system size,
and furthermore take the ``quench'' limit $T \to 0$.
In the quench limit, the wavefunction remains unchanged during the process.
On the other hand, the Hamiltonian changes due to the time dependence of the vector potential.
At the end of the process $t=T$, the system contains an AB flux of $2\pi$, which 
can be eliminated by the large gauge transformation
\begin{align}
    U_x=\exp\left(i\frac{2\pi}{L}\sum_j x_j n_j\right) ,
    \label{eq:defUx}
\end{align}
where $n_j$ was defined in Eq.~\eqref{eq:def_n}.
In order to compare the wavefunction before and after the flux insertion,
we must apply the large gauge transformation $H \to U_x^\dagger H U_x$
to get back to the original gauge.
Including the effect of the large gauge transformation,
the quantum state (wavefunction) of the system is 
\begin{align}
 U_x^\dagger | \Psi_0 (\vec{A}=0) \rangle .
\end{align}
In this paper, we are interested in insulators with an excitation gap, for which the polarization
density is well-defined.
In such systems, the sudden insertion of the AB flux, which is equivalent to the application of
a delta-function pulse of the electric field $E_x = - (2\pi/L)\delta(t) $, is expected to 
preserve the ground state. Namely, the post-quench state essentially remains the ground state.
Nevertheless, we expect the phase factor due to the dipolar energy~\eqref{eq:dipole}
$ \int_0^T dt\; E_x \calP_x = - 2 \pi \calP_x /L = - 2 \pi P_x$,
where $P_x \equiv \calP_x / L$ is the polarization density.
This implies 
    $\langle \Psi_0 (\vec{A}=0) | U_x^\dagger | \Psi_0 (\vec{A}=0) \rangle \propto e^{- 2\pi i P_x}$,
or equivalently
\begin{align}
    P_x = \frac{1}{2\pi} \arg{\left(  \langle \Psi_0 (\vec{A}=0) | U_x | \Psi_0 (\vec{A}=0) \rangle  \right) } .
    \label{eq:PxResta}
\end{align}
This is nothing but the Resta formula~\cite{Resta1998} for many-body polarization.

There are a few subtleties concerning this formula.
First, even though we are usually interested in the polarization at $\vec{A}=0$, the above flux insertion process
would give a certain average over $0 \leq A_x \leq 2\pi /L$. 
We can however expect that the dependence on the vector potential vanishes in the thermodynamic limit $L \to \infty$.
The more serious issue is the robustness of the ground state against the sudden insertion
of the AB flux.
In one dimension, the robustness is supported by the agreement~\cite{WatanabeOshikawa2018} between the Resta formula~\eqref{eq:PxResta} and
the Berry phase formula which corresponds to an adiabatic AB flux insertion.
However, in higher dimensions, the fidelity $|\langle \Psi_0 | U_x |\Psi_0 \rangle |$ is generally smaller than unity even in 
the thermodynamic limit, implying the significance of excitations due to the sudden AB flux insertion.
We may still hope that Eq.~\eqref{eq:PxResta} to be valid even in such cases, but it is still an open question.

As we mentioned in the Introduction, the subtlety of the Resta formula~\eqref{eq:PxResta} is related to the
fact that $U_x$ is not a conserved charge.
In order to resolve this issue (for inversion-symmetric systems), let us consider the following setup.
Instead of $\vec{A}=0$, we now choose the ground state at $\vec{A} = \vec{A}^P_0 = (-\pi/L) $ as the initial state.
This gauge field is written as $A_{j+\hat{x},j}=-\pi/L$ in the SSH model \eqref{eq:HA}.
Following the sudden insertion of the $2\pi$ AB flux, we apply the spatial inversion $M_x: x \to -x$.
After the process, the wavefunction is $\tilde{M}_x | \Psi_0 (\vec{A}^P_0) \rangle$, where 
\begin{align}
    \tilde{M}_x = M_x U_x .
    \label{eq:def_tildeMx}
\end{align}
The crucial observation is that $\tilde{M}_x$ commutes with the Hamiltonian $H(\vec{A}^P_0)$, since $U_x$ shifts
$\vec{A}^P_0 \to \vec{A}^P_1 = - \vec{A}^P_0$. 
As a consequence, the ground state $|\Psi(\vec{A}_0^P) \rangle$ should be an eigenstate of $\tilde{M}_x$ and
the eigenvalue of $\tilde{M}_x$ can be regarded as a quantum number.
Since $\tilde{M}_x$ is unitary, its eigenvalue is unimodular (phase factor).
The phase of eigenvalue of $\tilde{M}_x$ must contain the dipole energy contribution $2 \pi P_x$,
as in the Resta formula~\eqref{eq:PxResta}.
However, it also contains the response of the ground-state to the inversion operation $M_x$.
In order to subtract the latter effect, let us consider the ratio 
\begin{align}
  e^{2\pi i \Delta p} \equiv \frac{\langle \Psi_0(\vec{A}^P_0) | \tilde{M}_x| \Psi_0(\vec{A}^P_0) \rangle }{
    \langle \Psi_0(\vec{A}=0) | M_x| \Psi_0(\vec{A}=0) \rangle } .
\label{eq:def_Deltap}
\end{align}
Since the denominator in the right-hand side represents the inversion parity of the ground state, 
the ratio would give the information on the dipole moment.
It should be noted, however, that in the denominator we use the ground state at zero vector potential,
so that it is an eigenstate of the inversion $M_x$.
Although the ground-state response to the inversion could be different between $|\Psi_0(\vec{A}_0^P)\rangle$ and
$|\Psi_0(\vec{A}=0)\rangle$, we expect that they can be identified in the thermodynamic limit.
The advantage of the new formula for the polarization density $\Delta p$ is that both the numerator and
the denominator are quantum numbers (eigenvalues of operators).

We can extend this idea to quadrupole moment (density).
The electric field which couples to the quadrupole moment is generated by 
\begin{align}
    \vec{A} = \vec{A}^Q_0 + \frac{t}{T} \left( \vec{A}^Q_1 - \vec{A}^Q_0 \right),
\end{align}
such that $\vec{A}^Q_0 = (2\pi/L^2) (-y,0)$ and $\vec{A}^Q_1 = (2 \pi/L^2) (0,x)$.
To be precise,
we have introduced the vector potentials 
$A_{jk}$ in the gauge configuration shown in Fig.~\ref{fig:model} (c) for the BBH model~\cite{Hatsugai1999, Tada2021,YTMO}
corresponding to $\vec{A}^Q_0$.
This induces the desired electric field $\vec{E} \propto (y,x)$.
After the application of the delta-funciton pulse of the electric field by switching the
vector potential instantaneously from $\vec{A}^Q_0$ to $\vec{A}^Q_1$, we can perform the
gauge transformation by
\begin{align}
    U_{xy} = \exp{\left( i\frac{2\pi }{L^2} \sum_j x_j y_j n_j \right)} 
    \label{eq:def_Uxy}
\end{align}
to go back to the original gauge. 
While we may expect that the ground-state expectation value $U_{xy}$ gives the quadrupole moment (density),
which was in fact what was proposed in Refs.~\onlinecite{Wheeler2019,Kang2019}, we encounter
various issues~\cite{Ono2019}.
First, although the expression~\eqref{eq:def_Uxy} is applied to systems with periodic boundary conditions,
it lacks the periodicity.
In case of the dipole moment, $U_x$ defined in Eq.~\eqref{eq:defUx} is manifestly invariant under
the translation $x_j \to x_j + L$ because $n_j \in \mathbb{Z}$, and thus is consistent with the periodic boundary conditions.
However, Eq.~\eqref{eq:def_Uxy} lacks the invariance because of the factor $L^2$ in the denominator.
Furthermore, the expectation value of $U_{xy}$ shows a peculiar dependence on the choice of the origin,
although the physical quadrupole moment should not.

In these respects,
there are even more subtleties in
the many-body quadrupole moment defined by the expectation value of Eq.~\eqref{eq:def_Uxy}
than the dipole moment based on Eq.~\eqref{eq:defUx}.
Moreover, the subtleties in the dipole moment defined by $U_x$ is also inherited by the
quadrupole moment defined by $U_{xy}$.
Following Eqs.~\eqref{eq:def_tildeMx} and~\eqref{eq:def_Deltap}, we introduce 
a new operator to study the quadrupole moment in many-body systems
with $C_4$ discrete rotation symmetry, by combining the gauge transformation with the $\pi/2$ rotation as
\begin{align}
    \tilde{C}_4 = C_4 U_{xy} .
\end{align}
The composite operator $\tilde{C}_4$ also commutes with the Hamiltonian $H(\vec{A}^Q_0)$
with an appropriate definition.
In order to subtract the ground-state response to the $C_4$ rotation, we again divide the expectation value of
$\tilde{C}_4$ by the eigenvalue of $C_4$ for the ground state at $\vec{A}=0$,
\begin{align}
  e^{2\pi i \Delta q} \equiv \frac{\langle \Psi_0(\vec{A}^Q_0) | \tilde{C}_4| \Psi_0(\vec{A}^Q_0) \rangle }{
    \langle \Psi_0(\vec{A}=0) | C_4| \Psi_0(\vec{A}=0) \rangle } .
\label{eq:def_Deltaq}
\end{align}
As in the case of the dipole moment~\eqref{eq:def_Deltap}, the quadrupole moment is 
now defined in terms of \emph{quantum numbers}.
This also partially resolves the additional issues in the many-body quadrupole moment, such as the
origin dependence, as we will demonstrate later.

Summarizing both cases, we define the multipole indices as 
\begin{align}
\Delta r&=\tilde{r}-r,
\label{eq:Dr}
\end{align}
where $\tilde{r}=\tilde{p},\tilde{q}$ represents the eigenvalue $e^{2\pi i \tilde{r}}$ of the composite operator $\tilde{M}_x$ or
$\tilde{C}_4$ for the ground state of the Hamiltonian with the background vector potential $\vec{A}=\vec{A}^{P,Q}_0$,
and $r=p,q$ represents the eigenvalue $e^{2\pi i r}$ of $M_x$ or $C_4$
for the ground state of the reference Hamiltonian $H(\vec{A}=0)$.
Since ${\tilde{M}_x}^2 = {M_x}^2= {\tilde{C}_4}^4 = {\tilde{C}_4}^4 =1$, the indices are quantized as
$p, \tilde{p} , \Delta p \in \{ 0, 1/2 \}$ and $q, \tilde{q} , \Delta q \in \{ 0, 1/4, 1/2, 3/4 \}$
modulo 1 (see Appendix~\ref{app:operator}).
While the new indices admit a natural interpretation in terms of the response to an external electric field
similarly to the many-body multipole operator proposed in the past~\cite{Resta1998,Kang2019,Wheeler2019},
our proposal resolves various issues related to the fact that $U_x$ and $U_{xy}$ do not commute with the Hamiltonian 
and thus do not represent quantum numbers by themselves.
Besides, the new indices have several advantages thanks to their definitions based on the quantum numbers;
$\Delta r$ is robust to model parameters and it enables a rigorous analytic derivation of 
the bulk-boundary correspondence.
Details will be discussed in the remainder of this paper.

\section{Gauge transformation}
An advantage of
the combined symmetry operator $\tilde{C}_4$ is that it naturally has a periodicity $(x,y)\to (x\pm L,y\pm L)$
(any double sign)
thanks to gauge transformations, in contrast to $U_{xy}$ alone~\cite{Wheeler2019,Kang2019}. 
To see this, let us first consider the trivial symmetry $x\to x\pm L$ for dipoles from the view point of the gauge transformation.
We introduce a scalar function $f_j=\pm\pi$ for $0\leq x_j\leq (L-1)/2$ 
$(0\leq x_j\leq L/2-1)$ and 
$f_j=\mp\pi$ otherwise when $L$ is odd (even), and consider a gauge transformation $A\to A^f=A+df$ by
${\mathcal U}=\exp(i\sum_jf_jn_j)$.
The transformed operator is 
\begin{align}
\tilde{M}_x^f={\mathcal U}\tilde{M}_x{\mathcal U}^{-1}
=M_x\exp\left(i\frac{2\pi}{L}\sum_j(x_j\pm L)n_j\right),
\end{align}
which is a spatially translated version of $\tilde{M}_x$
and matches the periodicity of the system.
Similarly for quadrupoles,
the scalar function $f_j=\pm(2\pi/L)(x_j\pm y_j)$ for $0\leq x_j,y_j\leq L-1$ leads to 
\begin{align}
\tilde{C}_4^f={\mathcal U}\tilde{C}_4{\mathcal U}^{-1}
=C_4\exp\left(i\frac{2\pi}{L^2}\sum_j(x_j\pm L)(y_j\pm L)n_j\right),
\end{align}
(any double sign) .
This well matches the periodic boundary condition.
It should be noted that the index $\Delta r$ does not change under the gauge transformation
on both the symmetry operator and the wavefunction.
This is simply because $\tilde{g}\ket{\Psi}=e^{i2\pi \tilde{r}}\ket{\Psi}$ readily implies
$\tilde{g}^f\ket{\Psi^f}=e^{i2\pi \tilde{r}}\ket{\Psi^f}$
for $\tilde{g}=\tilde{M}_x,\tilde{C}_4$,
where $\ket{\Psi^f}={\mathcal U}\ket{\Psi}$ is the ground state of $H(A^f)={\mathcal U}H(A){\mathcal U}^{-1}$.
Therefore, our discussions based on $\Delta r$ work for the new gauge fields as well.

\section{Calculation of multipole index}
We can explicitly calculate the indices $\Delta r$ in the SSH model and BBH model to show
their effectiveness as a variant of order parameters.

\subsection{Calculation of $\Delta p$ for SSH model}
We first show that $\Delta p$ is non-zero in the topologically non-trivial phase of the 
SSH model and zero in the trivial state.
For this problem, we emphasize that our index $\Delta p$ does not change in presence of interactions as long as the many-body spectrum is gapped under the symmetry, because it is defined by the quantum numbers. Therefore, we can focus on the non-interacting limit $V_{jk}=0$ at half-filling ($\rho=1$) and it is sufficient to consider two limiting cases with either $t=0$ or $w=0$, which greatly simplifies the calculations.
Thus calculated results hold true for all the states adiabatically connected to the limit.

In the trivial phase with $w\neq0$, the ground state of $H(A)$ is adiabatically connected to that of $t=0$
which is independent of $A$, 
\begin{align}
\ket{\Psi_0(A)}
=\prod_j\frac{1}{\sqrt{2}}(c_{ja}^{\dagger}+c_{jb}^{\dagger})\ket{0}.
\end{align}
The fermions are localized at each site and the index is $\Delta p=0$ in this phase.
On the other hand, in the non-trivial phase with $t<0$, 
the ground state is smoothly connected to that of $w=0$,
\begin{align}
\ket{\Psi_0(A)}
=\prod_j\frac{1}{\sqrt{2}}(e^{-iA_x/2}c_{ja}^{\dagger}+e^{iA_x/2}c_{j+1,b}^{\dagger})\ket{0},
\end{align}
where $A_x=0,-\pi/L$.
In this case, the fermions are localized on each bond.
One can easily evaluate the eigenvalues of ${M}_x$ and $\tilde{M}_x$ for $A_x=0,-\pi/L$ respectively,
and obtain 
a non-zero value $\Delta p=1/2$.

\subsection{Calculation of $\Delta q$ for BBH model}
Similarly, we compute $\Delta q$ for the BBH model at half-filling ($\rho=2$) to find 
$\Delta q=0$ for the topologically trivial phase and $\Delta q=1/2$ for the non-trivial phase.
As in the SSH model, it is sufficient to focus on the non-interacting case.
The ground state wavefunction in the trivial phase in the limit $t=0,w\neq0$ is independent of $A_{jk}$, where
the fermions are localized at each site and correspondingly $\Delta q=0$.
On the other hand, for the topologically non-trivial phase,
the ground state wavefunction in the limit $t<0, w=0$ is
\begin{align}
\ket{\Psi_0(A)}
&=\prod_j\gamma_{j1}^{\dagger}\gamma_{j0}^{\dagger}\ket{0}, \\
\gamma_{(0,0),0}^{\dagger}
&=\frac{1}{2}(\omega^{1/4}c_{0a}^{\dagger}+c_{\hat{x},b}^{\dagger}
+\omega^{-1/4}c_{\hat{x}+\hat{y},c}^{\dagger}+\omega^{1/2}c_{\hat{y},d}^{\dagger}), \nonumber\\
\gamma_{(0,0),1}^{\dagger}
&=\frac{1}{2}(\omega^{1/4}c_{0a}^{\dagger}+ic_{\hat{x},b}^{\dagger}
-\omega^{-1/4}c_{\hat{x}+\hat{y},c}^{\dagger}-i\omega^{1/2}c_{\hat{y},d}^{\dagger}),\nonumber
\end{align}
where $\omega=\exp(i2\pi/L^2+i\theta_t)$, and $\theta_t\neq0$ is a model parameter which induces
a non-zero band gap at half-filling~\cite{Benalcazar2017Science,Benalcazar2017,Wheeler2019}
(see Appendix~\ref{app:model}).
The operators $\gamma_{jn}$ at general sites 
have structures similar to that of $\gamma_{(0,0),n}$.
In this state, the fermions are localized at each plaquette.
Then, one finds $\Delta q=1/2=2\times1/4$ which basically arises from the factors $\omega^{1/4}$ 
in $\gamma_{jn}$ (see Appendix~\ref{app:dq}).

\subsection{Application to strongly interacting systems}
\label{sec:interacting}
Our argument and indices are applicable equally to strongly interacting systems.
Here, we discuss SSH model and BBH model with sufficiently large interactions 
$H_{\rm int}=\sum_{jk}V_{jk}n_{j}n_{k}$ with the one-site translation symmetry $T$ 
in addition to the point group symmetry.
To be precise, the range of $V_{jk}$ is assumed to be within the nearest neighbor sites.
Although the on-site $V_0=V_{jj}$ will be larger than the inter-site $V_1=V_{jk} (|j-k|=1)$ in a realistic system,
we consider the opposite limit $V_{0}\ll V_{1}$ to demonstrate efficiency of our argument to
interacting systems in a simple manner. 
In this case, the ground state for the strong coupling limit $V_{1}\to \infty$ 
will be a charge-density-wave state.
In this phase,
the ground state wavefunctions for the SSH model (at half filling $\rho=1$ for an even system size $L$
under the periodic boundary condition) 
are adiabatically connected to
\begin{align}
\begin{aligned}
&\ket{\Psi_{\pm}}=\frac{1}{\sqrt{2}}\left( \ket{\Psi_1}\pm \ket{\Psi_2}\right), \\
&\ket{\Psi_1}=\prod_{j:{\rm even}}c_{ja}^{\dagger}c_{jb}^{\dagger}\ket{0},
\ket{\Psi_2}=\prod_{j:{\rm odd}}c_{ja}^{\dagger}c_{jb}^{\dagger}\ket{0}.
\end{aligned}
\label{eq:GSp_pm}
\end{align}
This holds true for both the zero vector potential $A_x=0$ and the non-zero vector potential $A_x=-\pi/L$.
The two wavefunctions $\ket{\Psi_{1,2}}$ break the translation symmetry of the Hamiltonian,
while $\ket{\Psi_{\pm}}$ are translationally symmetric and are eigenstates of the translation operator,
$T_x\ket{\Psi_{\pm}}=\pm \ket{\Psi_{\pm}}$. 
They correspond to the degenerate ground states of the charge-density-wave phase, which hold for both 
$|t|<|w|$ and $|t|>|w|$ as long as the interaction $V_1$ is sufficiently large.
Besides, $\ket{\Psi_{1,2}}$ are common eigenstates of $M_x$ and $U_x$.
Both of $\ket{\Psi_1}$ and $\ket{\Psi_2}$ (and thus $\ket{\Psi_{\pm}}$) have
the mirror eigenvalues $p=L/4$ (mod 1) for $M_x$ where $L$ is even,
because $M_x(c_{ja}^{\dagger}c_{jb}^{\dagger})M_x^{-1}=c_{L-j,b}^{\dagger}c_{L-j,a}^{\dagger}
=e^{i\pi}c_{L-j,a}^{\dagger}c_{L-j,b}^{\dagger}$ for each $j$
and there are $L/2$ such factors in Eq.~\eqref{eq:GSp_pm}. 
They also have the common eigenvalue $\tilde{p}=1/2+L/4$ for $\tilde{M}_x=M_xU_x$, 
because 
\begin{align}
U_x\ket{\Psi_{1,2}}&=
\exp\left( \pm i\frac{2\pi}{L}\sum_j (-1)^{x_j}x_j\right)\ket{\Psi_{1,2}}\nonumber\\
&=\exp(\mp i\pi)\ket{\Psi_{1,2}} = - \ket{\Psi_{1,2}} .
\end{align}
at the half-filling $\rho=1$.
(Because of our definition of the number operator~\eqref{eq:def_n}) used in $U_x$~\eqref{eq:defUx},
$n_j = \pm (-1)^{x_j}$ in the ideal charge density wave states $|\Psi_{1,2}\rangle$.) 
Therefore $\Delta p=1/2$ for both states (and thus for $\ket{\Psi_{\pm}}$).
We emphasize that the index $\Delta p=1/2$ is not limited to $V_{jk}\to \infty$ and
does not change in the entire charge-density-wave phase,
because it is denfined by the conserved quantum numbers.
The non-vanishing $\Delta p$ corresponds to the fact that 
the charge-density-wave states $\ket{\Psi_{1,2}}$ with broken translation symmetry
have non-zero polarization.
Indeed, the polarization is $P_x=(1/L)\sum_jx_jn_j=\pm1/2$ for $\ket{\Psi_{1,2}}$ under the open boundary condition.
This means that the index $\Delta p$ can describe the dipole moment not only in topologically non-trivial band insulators
but also in topologically trivial correlated insulators.
Note that $\Delta p$ is no longer a topological index for degenerate gapped states,
which is distinguished from the characterization of (uniquely gapped) symmetry protected topological states.

A similar argument applies to the BBH model (at half filling $\rho=2$) with $|t|,|w|\ll V$
and an even $L$.
The ground states show a staggered charge-density-wave order when $V_{0}\ll V_1$, 
because the square lattice is a bipartite lattice
with A, B-sublattices. 
The translationally symmetric states are adiabatically connected to 
\begin{align}
\begin{aligned}
&\ket{\Psi_{\pm}}=\frac{1}{\sqrt{2}}\left( \ket{\Psi_1}\pm \ket{\Psi_2}\right), \\
&\ket{\Psi_1}=\prod_{j:{\rm A-sublattice}}c_{ja}^{\dagger}c_{jb}^{\dagger}c_{jc}^{\dagger}c_{jd}^{\dagger}\ket{0},\\
&\ket{\Psi_2}=\prod_{j:{\rm B-sublattice}}c_{ja}^{\dagger}c_{jb}^{\dagger}c_{jc}^{\dagger}c_{jd}^{\dagger}\ket{0}.
\end{aligned}
\end{align}
Again, this holds true for both zero and non-zero vector potentials.
These two states are common eigenstates of $C_4$ and $U_{xy}$.
The $C_4$ eigenvalues for $\ket{\Psi_{1,2}}$ are $q=L^2/4 \equiv 0$ (because $L$ is assumed to be even)
similarly to $p$ in the SSH model. 
Furthermore, in the limiting charge-density wave states $\ket{\Psi_{1,2}}$,
the average particle number per unit cell is $\rho=2$ and
the particle number is given as
\begin{equation}
    n_j = \pm 2 (-1)^{x_j + y_j} .
\end{equation}
Thus
\begin{align}
U_{xy}\ket{\Psi_{1,2}}&=
\exp\left( \pm i\frac{4\pi}{L^2}\sum_j(-1)^{x_j+y_j}x_jy_j\right)\ket{\Psi_{1,2}}\nonumber\\
&= e^{\pm i \pi} \ket{\Psi_{1,2}} .
\end{align}
As a consequence, $\ket{\Psi_{1,2}}$, and thus their superpositions
$\ket{\Psi_\pm}$, belong to the eigenvalue $\tilde{q}= 1/2+L^2/4\equiv 1/2$
of the composite operator $\tilde{C}_4$.
Therefore
\begin{equation}
    \Delta q = \frac{1}{2} ,
\end{equation}
indicating that the charge-density-wave states belong to the phase with a nontrivial quadrupole index
which is distinct from the trivial phase under the $C_4$ symmetry.
This is consistent with the fact that the charge-density-wave states $\ket{\Psi_{1,2}}$
have non-zero quadrupole moments $Q_{xy}=(1/L^2)\sum_jx_jy_jn_j=\pm 1/2$ under the
open boundary condition. On the other hand, $\Delta q$ is well defined for the periodic boundary condition.
Furthermore, $\Delta q$ is invariant within each phase under the $C_4$ symmetry.
Therefore, similarly to $\Delta p$, the index $\Delta q$ can describe 
the quadrupole moments not only in topologically non-trivial band insulators
but also in topologically trivial correlated insulators.

\begin{figure}[tbh]
\includegraphics[width=8.0cm]{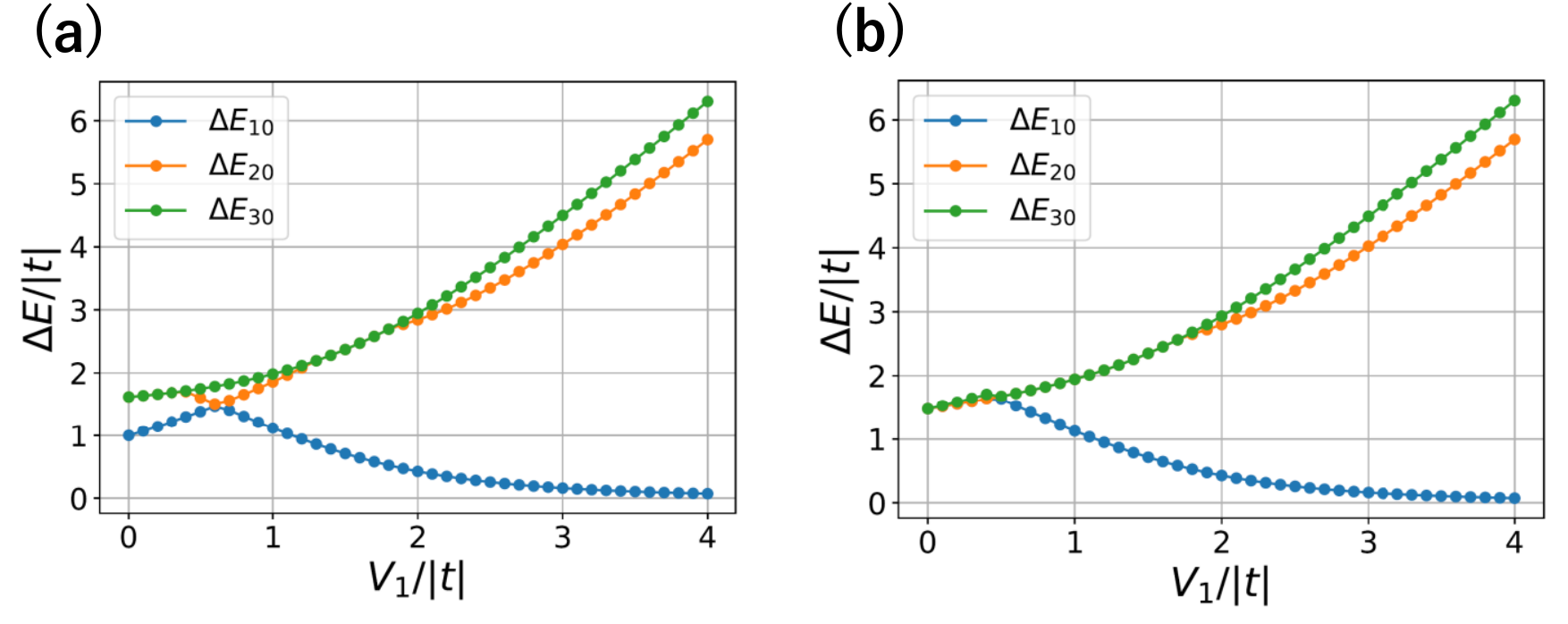}
\caption{Excitation energy $\Delta E_{mn}=E_m-E_n$ in the interacting SSH model 
for $t=-1.0, w=0.5t$ with $\rho=1, L=4$.
The vector potential is (a) $A_x=0$ and (b) $A_x=-\pi/L$.
}
\label{fig:dEVp1}
\end{figure}

\begin{figure}[tbh]
\includegraphics[width=8.0cm]{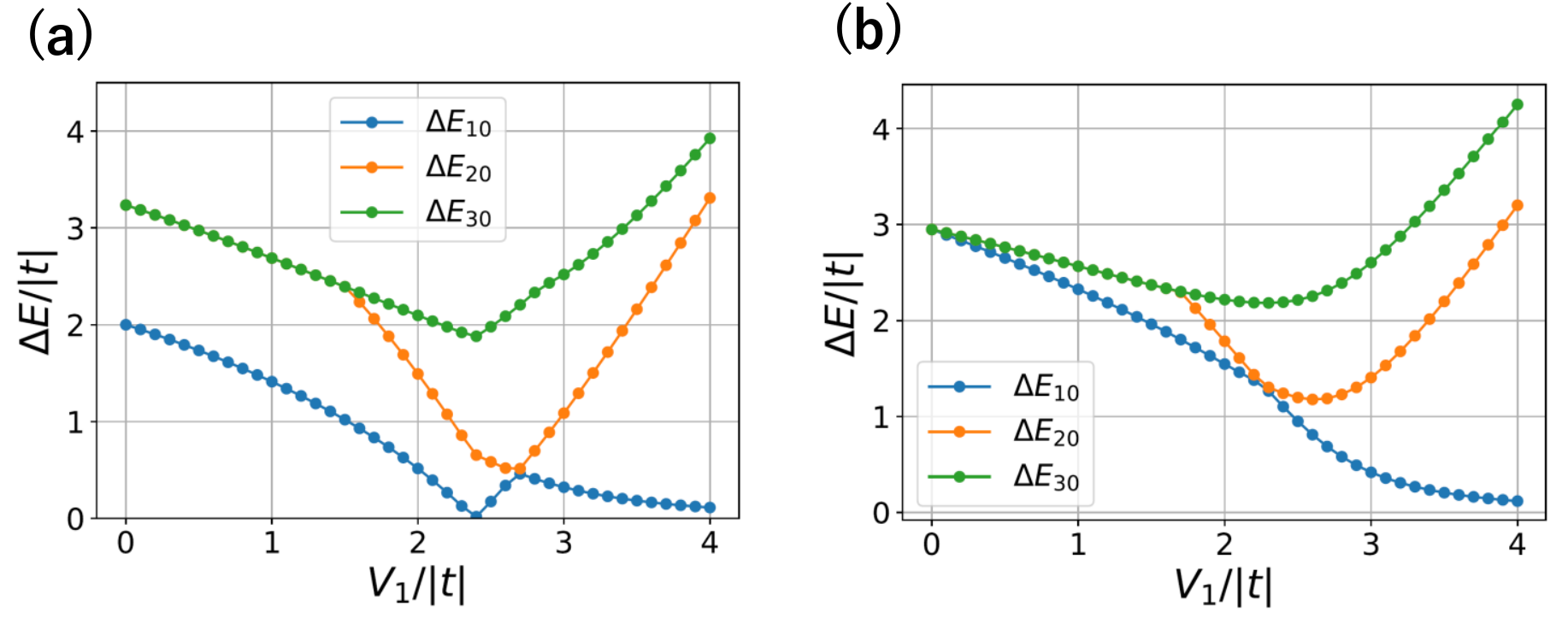}
\caption{Excitation energy $\Delta E_{mn}=E_m-E_n$ in the interacting SSH model 
for $t=-1.0, w=2t$ with $\rho=1, L=4$.
The vector potential is (a) $A_x=0$ and (b) $A_x=-\pi/L$.
}
\label{fig:dEVp0}
\end{figure}

As shown above, the multipole indices are non-trivial in the strong interacting regime $|t|,|w|\ll V$,
which is independent of the ratio $w/t$.
On the other hand, the ground states are topologically non-trivial for $|w|\lesssim |t|$ and trivial for $|w|\gtrsim |t|$ 
at the weakly interacting regime $|t|,|w|\gg V$.
There must be a quantum phase transition at $V_1=V_c$ between a band insulator for $V_1<V_c$
and the charge-density-wave ordered state for $V_1>V_c$,
and the many-body energy gap will close there in a thermodynamially large system.
(There might be multiple quantum phase transitions between $V=0$ and $V\to\infty$, but here we just
suppose that there is a single phase transition for simplicity.)
The phase transition is smeared in a finite size system, but the gap closing remains even for small $L$
when the level crossing takes place between two states with different quantum numbers.
Therefore, the topologically trivial band insulator with $\Delta p=0$
is separated by gap closing from the charge-density-wave state with $\Delta p=1/2$ for finite $L$. 
This can be demonstrated in the exact diagonalization of the interacting SSH model with $V_0=0$ and $V_1\neq 0$. 
In the numerical calculations, we choose a small system size $L=4$ so that 
each of the $n$-th energy level $E_n$ is clearly visible,
and we have checked that the results are qualitatively unchanged for larger $L$.
As shown in Fig.~\ref{fig:dEVp1} for $|w/t|<1$, 
the energy difference $\Delta E_{10}=E_1-E_0$ between the ground state and the first excited state
is non-zero due to finite size effects, and the dipole index $\Delta p=1/2$ does not change
for all $0\leq V_1<\infty$ in a finite size system.
$\Delta E_{10}$ will vanish in the thermodynamic limit corresponding to the (nearly) degenerate states
$\ket{\Psi_{\pm}}$ in Eq.~\eqref{eq:GSp_pm}.
On the other hand,
for for $|w/t|>1$ as seen in Fig.~\ref{fig:dEVp0}, 
the energy difference $\Delta E_{10}=E_1-E_0$ at $|w/t|>1$ becomes zero at a critical point $V_c(L)\simeq 2.4|t|$ 
when the vector potential is $A_x=0$, while the gap remains non-zero for $A_x=-\pi/L$.
Consequently, the ground state has $\Delta p=0-0=0$ for $V_1<V_c$ and $\Delta p=1/2-0=1/2$ for $V_1>V_c$.


\begin{figure}[tbh]
\includegraphics[width=8.0cm]{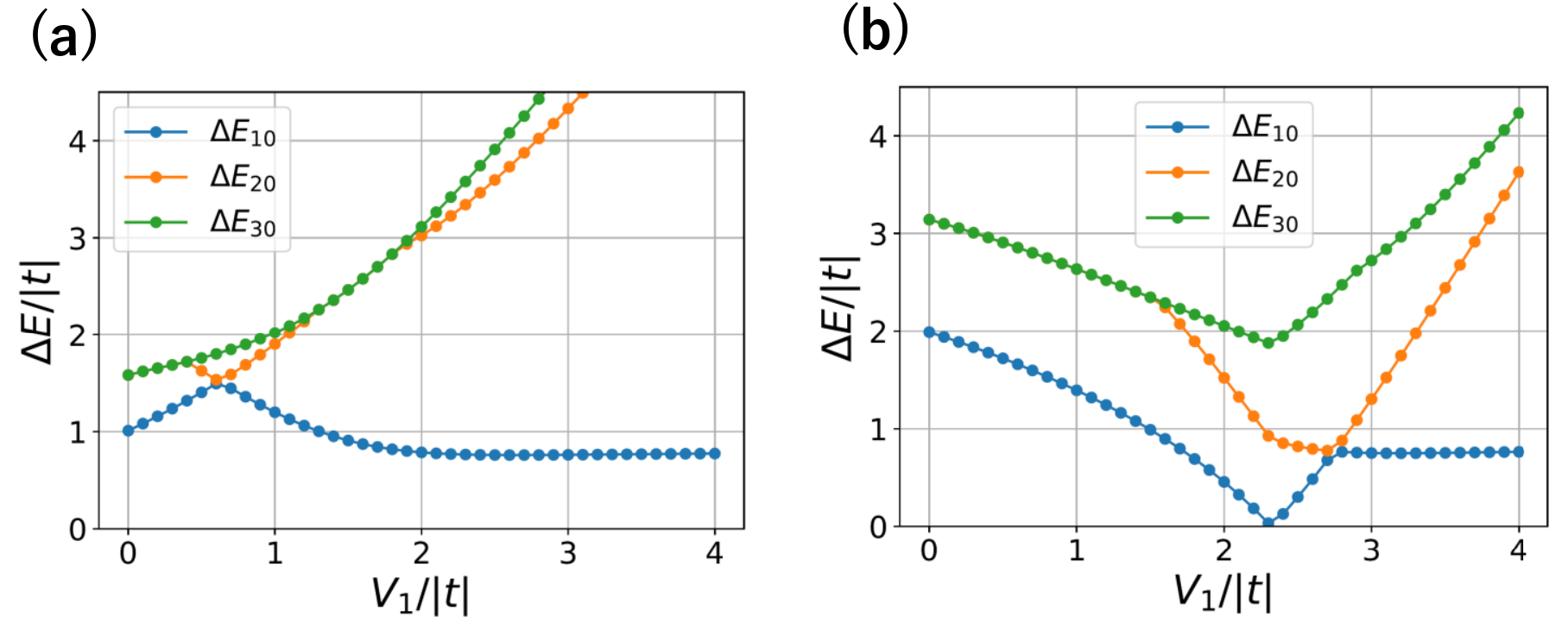}
\caption{Excitation energy $\Delta E_{mn}=E_m-E_n$ in the interacting SSH model 
for $t=-1.0,$ (a) $w=0.5t$ and (b) $w=2t$ with $\rho=1, L=4$.
The vector potential is $A_x=0$ and the staggered potential is $u_s=0.1|t|$.
}
\label{fig:dEVu}
\end{figure}

In absence of the additional translation symmetry of the Hamiltonian, the two states $\ket{\Psi_{1,2}}$
are no longer degenerate in general.
To be concrete, we introduce a staggered potential,
\begin{align}
H_{\rm stag}=\sum_j (-1)^ju_sn_j,
\end{align}
which favors one of $\ket{\Psi_1}$ or $\ket{\Psi_2}$.
Then, the ground state for $|w/t|<1$ is unique for all $V_1\geq 0$
with the index $\Delta p=1/2$, where 
$V_1=0$ and $V_1\to\infty$ are adiabatically connected each other without a phase transition 
as shown in Fig.~\ref{fig:dEVu} (a).
Therefore, the dipole band insulator and charge-density-wave state are essentially the same state when $u_s\neq0$.
On the other hand, in the topologically trivial case $|w/t|>1$,
the ground states for $V_1=0$ and $V_1\to\infty$ can still be well distinguished by $\Delta p$,
where $\Delta p=0$ for the former and $\Delta p=1/2$ for the latter. 
One can clearly see gap closing even in prensence of the staggered potential in Fig.~\ref{fig:dEVu} (b).
Note that the site-centered mirror symmetry is kept in both states and 
this phase transition is not related to spontaneous mirror symmetry breaking. 
(There is no bond-centered mirror symmetry in presence of the staggered potential.)
Numerical calculations suggest that the gap approaches zero at some $V\simeq V_c$ as $L$ increases
also in the 
$\pi$-flux system at $|w/t|>1$ (not shown),
which implies that there exists a phase transition irrespective of boundary conditions.
Therefore, the trivial band insulating state and charge-density-wave state are distinguishable
only by the site-centered mirror symmetry.
This means that there is no adiabatic path connecting
the trivial band insulator and non-trivial band insulator
even in an enlarged Hamiltonian space with
$H_{\rm int}$ and $H_{\rm stag}$ under the mirror symmetry.
Similar arguments may apply to quadrupole insulators.

\section{Bulk-boundary correspondence}

By using the combined operators,
we can naturally describe a bulk-boundary correspondence for 
interacting multipole insulators
with the point group symmetry and particle number U(1) symmetry,
which is a many-body generalization of 
the previous studies~\cite{Benalcazar2017,Langbehn2017,Song2017,Khalaf2018,Trifunovic2019,Takahashi2020,Teo2008}.
We first formulate our bulk-boundary correspondence focusing on interacting band insulators
where the ground state is uniquely gapped.
That is, we will show that a nontrivial index $\Delta r \neq 0$ requires that a gap closing must take place
when the boundary condition is deformed from the periodic to open.
Then, relations to the filling anomaly~\cite{Benalcazar2019} are discussed.
Furthermore, bulk-boundary correspondence is confirmed by numerical calculations.
We emphasize that our argument holds in presence of interactions and is applicable not only to a band insulator
but also to a correlated insulator whose energy gap is driven by interactions.

\subsection{Statement and proof}

In this paper, we are interested in gapped insulators.
Robustness of the many-body excitation gap
is widely accepted (and often assumed) although not mathematically proven in general.
One of the aspects of the robustness is the robustness against the insertion
of the AB flux~\cite{Oshikawa-Comme2000,Watanabe2018}.
That is, the excitation gap is expected not to close for any finite AB flux.
Another is the robustness of the many-body excitation gap against a cut
in a trivially gapped phase.
That is, if the system is in a trivial phase, the gap is expected to remain
non-zero when the system is cut, and there appears no edge/surface states.
In contrast, gapless edge states often appear in topological phases.
This is a typical manifestation of bulk-boundary correspondence.

Indeed, here we show that, a nontrivial multipole index $\Delta r \neq 0$ implies the existence of edge states.
More concretely, we prove the following statement
\begin{claim}
If the multipole index is non-trivial, $\Delta r\neq0$, under the periodic boundary condition,
gap closing takes place in the many-body energy spectrum of either $H(A=0)$ or $H(A\neq 0)$
when the periodic boundary condition is continuously tuned to the open boundary condition.
\end{claim}
The precise meaning of ``tuning the boundary condition" will be explained later.

\begin{figure}[tbh]
\includegraphics[width=7.0cm]{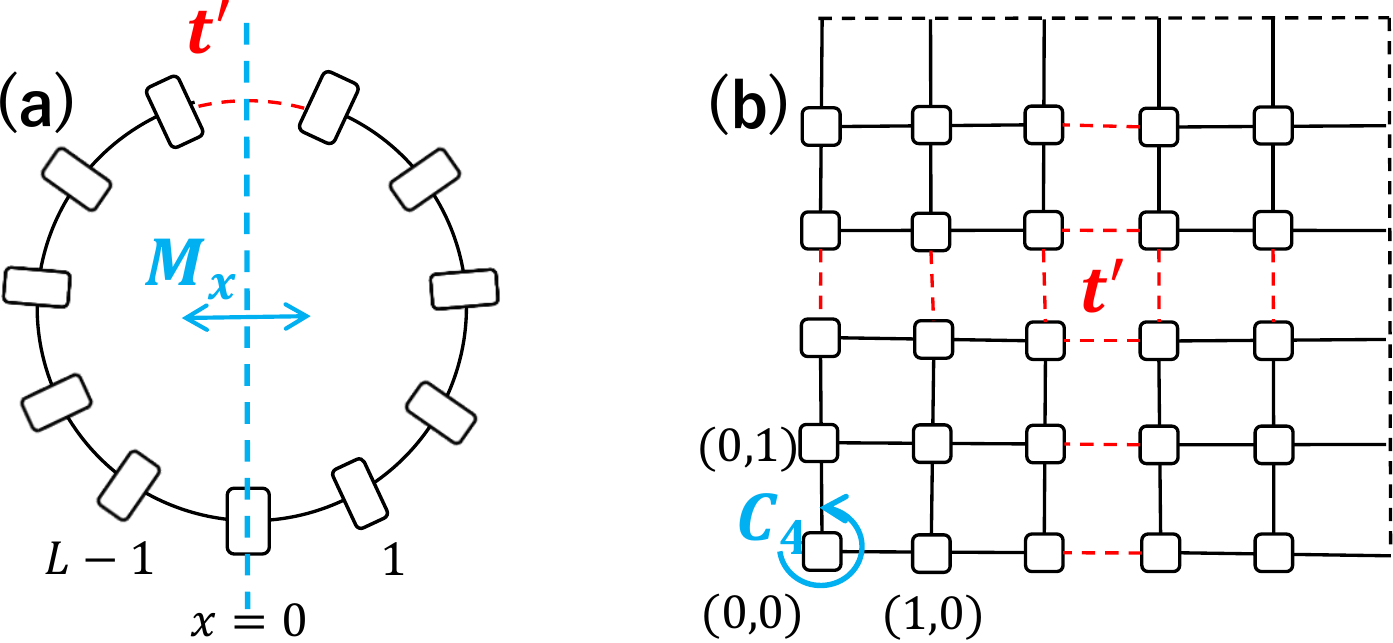}
\caption{(a) SSH model and (b) BBH model with $t'=\lambda t$ on the bonds with red colored broken lines. 
$\lambda=0$ corresponds to an open chain with the center $x=0$ 
and an open square with the center $(0,0)$, respectively.
If there are other hoppings and inter-site interactions, they are scaled by $\lambda$ in a similar manner.
}
\label{fig:OBC}
\end{figure}

We illustrate our argument using the examples of SSH and BBH models, although it is naturally applicable to more general models.
First, let us consider dipole insulators by using the SSH model under the periodic boundary condition as shown in 
Fig.~\ref{fig:OBC} (a).
The system size $L$ is now assumed to be odd, so that the corresponding system with the open boundary condition
also has site-centered mirror symmetry.
Then we can define the index $\Delta p$ as in Eq.~\eqref{eq:def_Deltap}.
We will show that a gap closing must take place during the ``cut'', namely when the boundary condition is modified
from periodic to open, if $\Delta p = \frac{1}{2}$.

The cut is implemented by changing the hopping integral $t'=\lambda t$ between the sites $x_j=(L-1)/2$ and $(L+1)/2$,
while other hopping integrals are fixed to $t$.
$\lambda=1$ corresponds to the periodic boundary condition and $\lambda=0$ 
does to the open boundary condition.
In presence of other hoppings and inter-site interactions, 
they are scaled by the parameter $\lambda$ in a similar manner.
Furthermore, we introduce the AB flux parametrized by $s$ as  $A_x(s)=-s\pi/L$.
Therefore we consider a family of Hamiltonians in the two-dimensional parameter space $(\lambda,s)$.

Let us define the operator  $U_x'(s)=\exp(is2\pi/L\sum_jX_jn_j)$, where $0 \leq s \leq 1$.
We have introduced the coordinate $X_j=x_j$ for $0\leq x_j\leq (L-1)/2$ and $X_j=x_j-L$ for $(L+1)/2\leq x_j\leq L-1$.
In general, $U_x'(s)$ introduces a twisted boundary condition and thus can be used to define
a symmetry of the system, only at $s=0, 1$.
In other words, $U_x'(s)$ corresponds to insertion of $s$ flux quantum as the AB flux and cannot be related to
the large gauge invariance except for $s=0,1$.
However, for the open boundary condition $\lambda=0$,
the system is completely insensitive to the AB flux,
as there is no path encircling the AB flux.
Equivalently, the vector potential $A_x(s)$
can be eliminated by the gauge transformation $U_x'(s)$
for any $0 \leq s \leq 1$, 
(Since there is no hopping term at the boundary for the open boundary condition $\lambda=0$,
the twist introduced by $U_x'(s)$ can be ignored.)
Thus the Hamiltonian on the lines $(\lambda,s=0,1)$ and $(\lambda=0,s)$ is invariant under
\begin{equation}
    \tilde{M}'_x(s) \equiv M_x U'_x(s) .
\end{equation}
We can then define $\tilde{p}'(\lambda,s)$ by the eigenvalue
$e^{2\pi i \tilde{p}'(\lambda,s)}$ of $\tilde{M'}_x(s)$ for the ground state under the boundary condition $\lambda$.
This eigenvalue is quantized as $\tilde{p}'(\lambda,s)=0,1/2$ and can change only when gap closing occurs, 
because $(\tilde{M}'(s))^2=1$ holds for $0\leq s\leq 1$
(Appendix~\ref{app:operator}).

\begin{figure}[tbh]
\includegraphics[width=4.0cm]{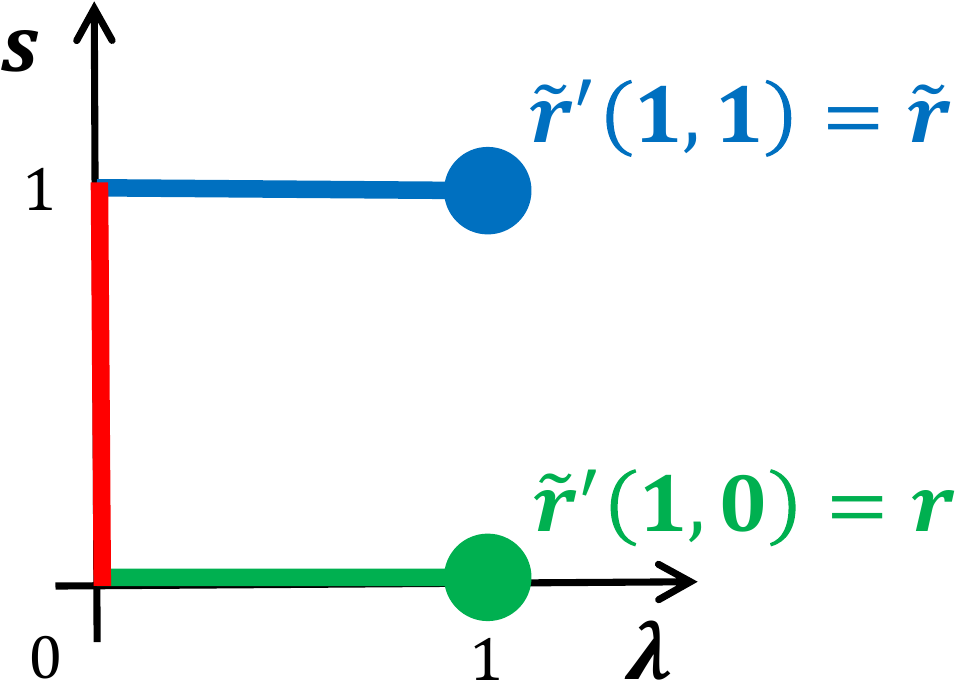}
\caption{The $(\lambda,s)$-plane and lines connecting the two points 
$(\lambda,s)=(1,0),(1,1)$, where $\lambda$ and $s$ characterize the boundary condition and the flux, respectively. 
$\tilde{r}'(\lambda,s)$ corresponds to the eigenvalue of $\tilde{M}_x'(s)$ or $\tilde{C}_4'(s)$
in the ground state with the boundary condition $\lambda$.
These operators commute with the Hamiltonian on the colored segments, but not on other regions.
}
\label{fig:lams}
\end{figure}

Now let us connect the two points $(\lambda,s)=(1,0)$ and $(1,1)$ 
along the lines $(0 \leq \lambda \leq 1, s=1)$, $(\lambda=0, 0 \leq s \leq 1)$,
and $(0 \leq \lambda \leq 1, s=0)$, as shown in Fig.~\ref{fig:lams}.
On the second segment $(\lambda=0, 0 \leq s \leq 1)$ represented by the red line in Fig.~\ref{fig:lams}, 
the Hamiltonian is always gauge equivalent.
Thus the eigenvalue $\tilde{p}'(\lambda=0,s)$ of the symmetry generator remains unchanged.
Furthermore, along the first and third segment, $\tilde{M}'_x(s)$ remains the exact symmetry.
Therefore, if no gap closings take place along $0 \leq \lambda \leq 1$ at both $s=0$ (green line) and
$s=1$ (blue line),
$\tilde{p}'(\lambda,s)$ remains unchanged.
Therefore, under this assumption, $\tilde{p}'(\lambda=1,s=1) = \tilde{p}'(\lambda=1,s=0)$.
On the other hand,
by definition, $\tilde{M}'_x(s=1) = \tilde{M}_x$ and thus $\tilde{p}'(\lambda=1,s=1) = \tilde{p}$.
Similarly, $\tilde{M}'_x(s=0) = M_x$ and thus $\tilde{p}'(\lambda=1,s=0) = p$.
Thus the assumption of no gap closing implies $\tilde{p} = p$ and thus the dipole index is trivial: $\Delta p=0$.
As a contraposition, if $\Delta p \neq 0$, there must be a gap closing along either
the first or third segments $(0 \leq \lambda \leq 1, s=0,1)$.
This signals the presence of gapless edge states.

We note that, in a finite-size system, the gapless edge states (ground-state degeneracy)
does not necessarily appear exactly at $\lambda=0$. 
Nevertheless, the above argument implies that the gap closing must take place
at a critical value $\lambda_c(L)\in [0,1]$ depending on the system size $L$,
which we confirm numerically later.
In the thermodynamic limit $L \to \infty$, $\lambda_c(L) \to 0$ is expected, corresponding to the gapless edge states
for the open boundary conditions.
This can be also interpreted as the existence of filling anomaly
when $\Delta p \neq 0$~\cite{Benalcazar2019} as will be discussed later.


A similar argument applies to quadrupole insulators with an odd linear system size $L$.
In this case, our argument is based on the spectral robustness against the flux in each plaquette,
where the flux $2\pi/L^2=O(L^{-2})$ is so small that
the spectra for $H(0)$ and $H(A)$ will be essentially same~\cite{Tada2021}.
The cut is implemented by the hopping $t'=\lambda t$ for the bonds between $((L-1)/2,y_j)$ and $((L+1)/2,y_j)$,
and $(x_j, (L-1)/2)$ and $(x_j, (L+1)/2)$ as shown in Fig.\ref{fig:OBC} (b), for which
the open boundary condition is realized at $\lambda=0$.
If there exist other hoppings and inter-site interactions, they are scaled by $\lambda$ in a similar manner.
Then, it is convenient to introduce the new coordinate $X_j, Y_j\in\{-(L-1)/2,\cdots,0,\cdots,(L-1)/2\}$ 
similarly to dipole insulators.
Accordingly, we make a gauge transformation by ${\mathcal U}=e^{i\sum_jf_jn_j}$ with 
$f_j=0$ for $Y_j\geq0$ and $f_j=(2\pi/L)X_j$ for $Y_j<0$.
The combined operator is transformed to $\tilde{C}_4=C_4\exp(i2\pi /L^2\sum_{j}X_jY_jn_j)$
and commutes with the Hamiltonian for $0\leq \lambda\leq 1$.

Now we introduce the operator $\tilde{C}_4'(s)=C_4U'_{xy}(s)$ with
$U'_{xy}(s)=\exp(i2\pi s/L^2\sum_{j}X_jY_jn_j)$ which commutes with the Hamiltonian
with the open boundary condition $\lambda=0$,
in the new gauge:
$[\tilde{C}_4'(s),H(A(s),\lambda=0)]=0$.
It is straightforward to see $(\tilde{C}_4'(s))^4=1$ for $0\leq s\leq1$ and its eigenvalues
are quantized (Appendix ~\ref{app:operator}).
Again, for the periodic boundary condition $\lambda=1$, the eigenvalue of $\tilde{C}_4'(s)$
is given by $q$ at $s=0$ and by $\tilde{q}$ at $s=1$.
Given these definitions and properties, we can repeat the same argument as before.
That is, if there is no gap closing while tuning the boundary condition along $0 \leq \lambda \leq 1$,
the quadrupole index $\Delta q =0$.
This implies that there must be a gap closing for $\lambda = \lambda_c \in [0,1]$.

We
note that the square geometry with corners for the open boundary condition is crucial in the above discussion, which is consistent with corner modes in a quadrupole insulator.
The above argument does not apply to a cylinderical system,
where $t'$ is introduced only in one of the $x$- or $y$-direction, because the $C_4$-rotation symmetry
is explicitly broken in such a case.
This would suggest that the gapless modes appear at corners of the system but not at edges,
although their spatial positions cannot be identified in our argument for the bulk-boundary correspondence.

In the above discussion,
we have used the property that the small flux $2\pi/L^2$ does not close an energy gap
and the spectra for $H(0)$ and $H(A)$ are essentially same in two-dimensions.
This can be proved when $H(0)$ has no flux~\cite{Tada2021}, 
but the gap might close otherwise because the total flux
in the entire system is $\sum_j2\pi/L^2=2\pi=O(1)$ which is comparable with the preassumed gap $O(1)$.
Although the argument in the previous study~\cite{Tada2021} cannot be directly applied to the BBH model with
 a flux parameter 
$\theta\neq0$ which break the time-reversal symmetry, 
the quadrupole phase is stable~\cite{Wheeler2019} for an extended region of $\theta$ and
the energy gap should not close when the tiny external flux $2\pi/L^2$ is added,
which will be true even in presence of interactions~\cite{Hastings2019,Koma2020}.
Therefore, our argument on the bulk-boundary correspondence should work for the BBH model with $\theta\neq0$.

\subsection{Relation to filling anomaly}
\label{sec:filling_anomaly}
Let us discuss the gap closing at $\lambda\simeq 0$ in more detail based on filling and symmetry 
of the wavefunctions~\cite{Benalcazar2019}.
In our setup, we focus on the particle number fixed sector with $N_e$ electrons 
for a system with $N_a$ atomic sites.
For a fixed system size $L$, the electron number is $N_e=\rho N_a=L$ for the half-filled SSH model
and $N_e=\rho N_a=2L^2$ for the half-filled BBH model.
In the following, we focus on the non-interacting ($V=0$) SSH model just for simplicity and 
similar arguments apply to the BBH model as well. 
The concluding statement holds also for interacting systems,
because our analytical proof in the previous section is applicable to such systems. 
Under the open boundary condition ($t'=0$),
the ground state wavefunction is a superposition of the state $\ket{\Psi_L}$ with an excess electron 
charge on the left edge and the state $\ket{\Psi_R}$ with an excess electron charge on the right edge. 
In the limit $w=0$, they are explicitly given by
\begin{align}
\begin{aligned}
\ket{\Psi_L}&=c^{\dagger}_{l,b}\prod_{j\neq r}\frac{1}{\sqrt{2}}(c^{\dagger}_{ja}+c^{\dagger}_{j+1,b})\ket{0}, \\
\ket{\Psi_R}&=c^{\dagger}_{r,a}\prod_{j\neq r}\frac{1}{\sqrt{2}}(c^{\dagger}_{ja}+c^{\dagger}_{j+1,b})\ket{0}
\end{aligned}
\end{align}
for $t<0$, where $l=(L+1)/2, r=(L-1)/2$ are the sites corresponding to the left edge and right edge, respectively.
These states are mirror symmetry broken states, and 
the charge localized at the left edge site is $\langle\Psi_{L}|n_l|\Psi_{L}\rangle
=+1/2$ compared to the average charge density $\rho=1$ and also $\langle\Psi_{L}|n_{r}|\Psi_{L}\rangle
=-1/2$.
Similarly, $\langle\Psi_{R}|n_l|\Psi_{R}\rangle=-\langle\Psi_{R}|n_{r}|\Psi_{R}\rangle=-1/2$.
(The total charge is neutral by definition.)
On the other hand, the two ground states 
\begin{align}
\begin{aligned}
\ket{\Psi_+}&=\frac{1}{\sqrt{2}}(\ket{\Psi_L}+\ket{\Psi_R}), \\ 
\ket{\Psi_-}&=\frac{1}{\sqrt{2}}(\ket{\Psi_L}-\ket{\Psi_R}) 
\end{aligned}
\end{align}
are mirror symmetric with different mirror-eigenvalues, and there is no charge accumulation at the edges, 
$\langle\Psi_{\pm}|n_l|\Psi_{\pm}\rangle=\langle\Psi_{\pm}|n_{r}|\Psi_{\pm}\rangle=0$.
(The same notation $\ket{\Psi_{\pm}}$ as those in Sec.~\ref{sec:interacting} 
is used here for simplicity, but they are different states.)
These ground states are
exactly degenerate at $w=0$.
Away from the $w=0$ limit, each of $\ket{\Psi_{L,R}}$ aquires correction terms,
and they will hybridize to obtain an energy separation which is exponentially small in the system size $L$,
which also give an energy gap for $\ket{\Psi_{\pm}}$.
This corresponds to the energy gap at $\lambda=0$ in the numerical calculations in Fig.~\ref{fig:gap}.
The finite size gap will vanish in the thermodynamic limit $L\to\infty$, because distance between the opposite
edges becomes infinitely large.

The gap closing is robust to perturbations which keep the point group symmetry.
Indeed, one can add a perturbation to the SSH model which breaks the on-site chiral symmetry but keeps
the mirror symmetry,
as generally discussed for point group symmetry protected topological phases
~\cite{Song2017pgSPT,Huang2017,Cheng2022}.
For example, we consider the potential term
\begin{align}
H_{\rm pert}=\sum_{j\mu} u_j^{\mu}c^{\dagger}_{j\mu}c_{j\mu},
\end{align}
where $u_{j}^a=u_{L-j}^b$ by the mirror symmetry.
The two states $\ket{\Psi_{\pm}}$ are still degenerate, although the single-particle edge modes 
aquire a non-zero energy due to the lack of the chiral symmetry.
This means that the gaplessness (or degenerate ground states) at the charge neutral filling 
are protected only by the mirror symmetry, which can be regarded as a variant of 
the filling anomaly~\cite{Benalcazar2019}.
In this context, our bulk-boundary correspondence is a filling anomaly type statement
and can be rephrased as follows.
\begin{claim}
Consider a system with a charge neutrality filling, point group symmetry, and 
a non-trivial index $\Delta r$ in the uniquely gapped ground state under the periodic boundary condition.
Then, it is impossible to realize a uniquely gapped ground state under the open boundary condition
with keeping the same filling and symmetry.
\end{claim}
The resulting ground state(s) under the open boundary condition must be either gapless or break the symmetry
in the thermodynamic limit.
We stress that
the above statement is valid for interacting systems as well, 
because our analytical proof in the previous section
is applicable also to such systems. 
This becomes important for understanding systems with strong interactions as will be discussed
in the next section.

\subsection{Numerical confirmation of bulk-boundary correspondence}
We numerically confirm the bulk-boundary correspondence.
To this end,
we first show numerical calculations of single-particle spectra for the 
non-interacting SSH and BBH models, where the boundary conditions are tuned by the hopping parameter 
$t'=\lambda t$
on the specific bonds (Fig.~\ref{fig:OBC}).
$t'=t$ corresponds to the periodic boundary condition and $t'=0$ describes the open boundary condition.
In case of a dipole insulator, we can also consider
$t'=-t$ corresponding to the anti-periodic boundary condition 
which is equivalent to the periodic boundary condition with a $\pi$-flux.

\begin{figure}[tbh]
\includegraphics[width=8.0cm]{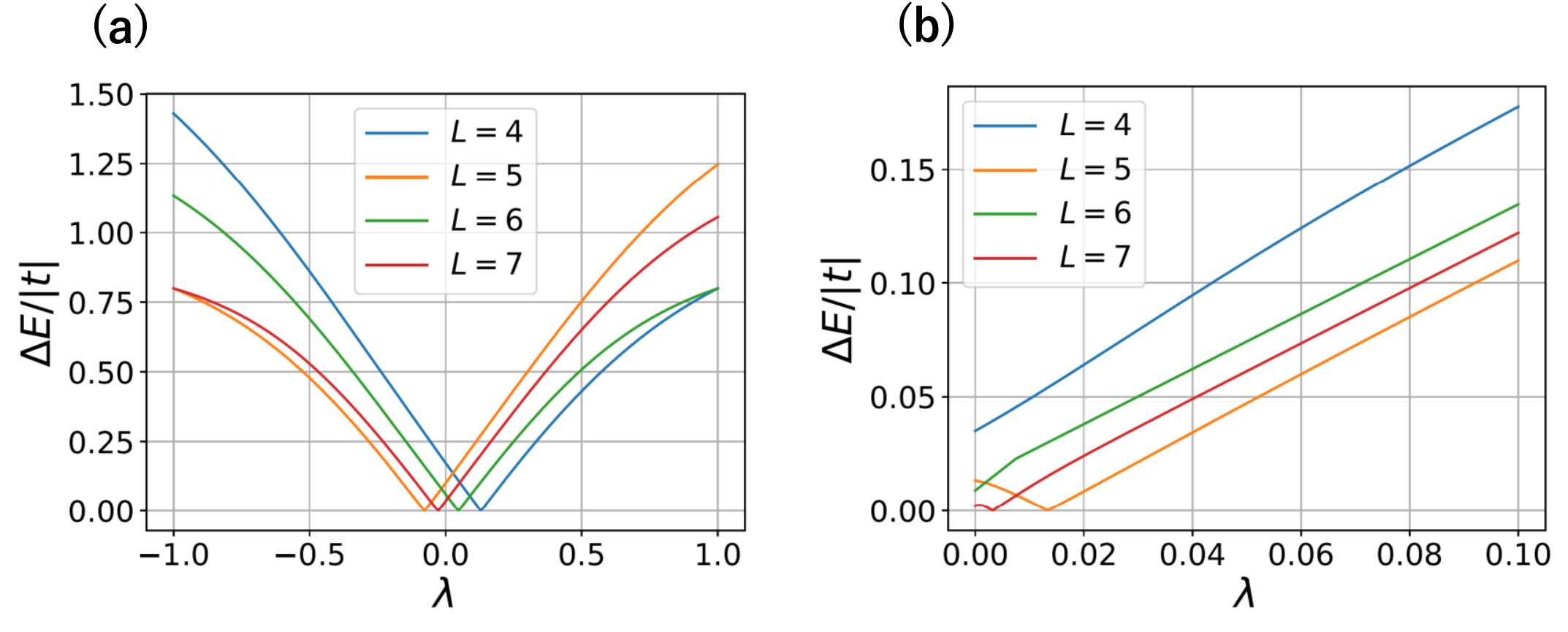}
\caption{The energy gap $\Delta E$ as a function of $\lambda$ for (a) SSH model with $t=-1, w=0.6t$
and (b) BBH model with $t=-1,w=0.3t, \theta_t=\theta_w=2\pi\times (5/16,8/25,12/36,16/49)$ for 
$L=(4,5,6,7)$, respectively. These fluxes correspond to $\theta =2\pi/3$ in the thermodynamic limit.
}
\label{fig:gap}
\end{figure}

We consider the SSH model for both even and odd system sizes $L$. 
Although our proof is not applicable for an even $L$, we naturally expect that gap closing takes place
in this case as well similarly to the case of an odd $L$. 
As examplified in Fig.~\ref{fig:gap}~(a),
the energy gap $\Delta E$ at half-filling (gap between the $L$-th and $(L+1)$-th single-particle energy levels)
is $\Delta E\sim t$ when $t'=t$ and it decreases as $t'$ is varied.
There is a small energy gap due to hybridization of the edge modes localized at 
opposite ends for a finite $L$ when $t'=0$.
One can see that $\Delta E=0$ at a critical strength of the hopping parameter 
$\lambda_c(L)$ depending on the system size.
This is fully consistent with our proof of the bulk-boundary correspondence,
where it is shown that there exists gap-closing when $t'$ is tuned to zero if $\Delta p\neq0$.
The gap closing takes place for any $L$ and the critical value $\lambda_c(L\to\infty)$ numerically approaches zero 
in the thermodynamic limit as physically expected, 
although we cannot rigorously prove $\lambda_c(L\to\infty)=0$.

The BBH model exhibits similar behaviors.
The energy gap $\Delta E$ at half-filling (gap between the $2L^2$-th and $(2L^2+1)$-th single-particle energy levels)
is shown in Fig.~\ref{fig:gap}~(b). 
The energy gap vanishes at a critical $\lambda_c(L)$ and it approaches zero in the thermodynamic limit similarly
to the SSH model.
Note that
the energy gap $\Delta E$ does not close for $L=4,6$ to which our proof for an odd $L$ is not applicable,
but $\Delta E$ at $\lambda=0$ approaches zero as $L$ increases
and $\Delta E$ for an odd $L$ and an even $L$ will converge to a same value in the thermodynamic limit.

\begin{figure}[tbh]
\includegraphics[width=8.0cm]{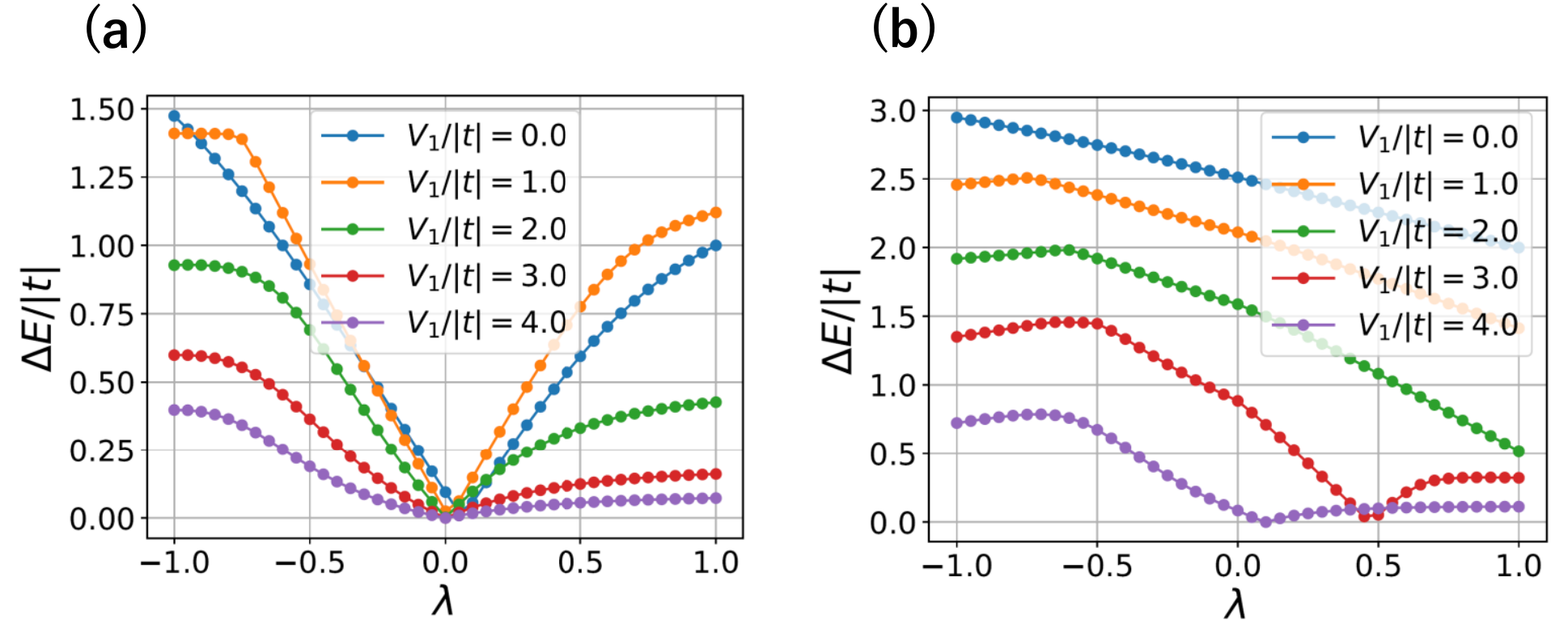}
\caption{The energy gap $\Delta E$ between the ground state and the first excited state
as a function of $\lambda$ in the interacting SSH model with the system size $L=4$.
The parameters are (a) $t=-1, w=0.5t$
and (b) $t=-1,w=2t$. 
}
\label{fig:dE_lam}
\end{figure}

The bulk-boundary correspondence holds also for strongly interacting systems as well,
where energy gaps are driven by the interactions.
Here, we consider the interacting SSH model with the inter-site interaction $V_1$ at half-filling $\rho=1$
(see also Sec.~\ref{sec:interacting}),
where hopping $t$ and $V_1$ are scaled as $t'=\lambda t, V_1'=|\lambda|V_1$ at a bond by the parameter
$-1\leq \lambda\leq 1$.
The system size is taken to be $L=4$ since an odd $L$ is incompatible with the charge-density-wave order, 
although our proof of the bulk-boundary correspondence is not applicable
to a system with an even $L$.
As in the previous section,
each of the $n$-th energy levels $E_n$ is clearly visible for $L=4$ and we have confirmed that qualitative
behaviors do not change for larger $L$. 
We naturally expect that the energy spectra for an even $L$ and an odd $L$ will converge to a same spectrum
in the thernodynamic limit.
As shown in Fig.~\ref{fig:dE_lam}, the energy gap closes around $\lambda=0$ for $|w/t|<1$ for any $V_1$,
because the dipole index is $\Delta p=1/2$ as was discussed in Sec.~\ref{sec:interacting}.
On the other hand, the gap closing takes place only for $V_1>V_c(L)$ when $|w/t|>1$,
because the index is $\Delta p=0$ for $V_1<V_c$ and $\Delta p=1/2$ for $V_1>V_c$.
The gap closing in the charge-density-wave ordered states is not related to single-particle edge modes
and is understood based on the filling anomaly discussed in the previous section (Sec.~\ref{sec:filling_anomaly}).

\section{Summary and discussion}
We have proposed the new indices $\Delta p, \Delta q$ for interacting dipole and quadrupole insulators
under the periodic boundary condition.
In presence of point group symmetries, these indices are quantized and they can 
well characterize multipole insulators.
There are several advantages of our multipole indices;
(i) they are well-defined for general dimensions for thermodynamically large systems
in contrast to the previously proposed ones, and
(ii) they do not change in a uniquely gapped phase in presence of the particle number U(1) and point group symmetries.
In addition, 
(iii) they are compatible with the periodicity of the system thanks to gauge transformations, and 
(iv) they can be extended to systems with open boundaries, which leads to the bulk-boundary correspondence.
Our indices are applicable to bosonic particle systems and spin systems as well, and hence
can be widely used for characterization of topological phases with point group symmetries.
Our approach may be extended to general
$C_n$-rotation symmetric quadrupole insulators and also
octupole insulators in three dimensions. These are left
for a future study.

\begin{acknowledgments}
We thank Maissam Barkeshli, Ken Shiozaki, and Yuan Yao for valuable comments and discussions.
This work is supported by JSPS KAKENHI Grant No. 17K14333, No. 22K03513, and No. 19H01808,
and by JST CREST Grant No. JPMJCR19T2.
\end{acknowledgments}

\appendix
\section{Definition of SSH model and BBH model}
\label{app:model}
The SSH model in the present study is a two-orbital spinless fermion model.
The inter-site hopping $t^{\mu\nu}_{jk}$ and intra-site hybridization $w^{\mu\nu}_j$ are
\begin{align}
t_{j+\hat{x},j}^{\mu\nu}&=
\begin{pmatrix}
0 & 0 \\
t & 0 \\
\end{pmatrix},\quad
w_{j}^{\mu\nu}=
\begin{pmatrix}
0 & w \\
w & 0 \\
\end{pmatrix}.
\end{align}
The mirror operation about the origin $x=0$ in absence of a vector potential is
$M_xc_{ja}M_x^{-1}=c_{L-j,b}$ and $M_xc_{jb}M_x^{-1}=c_{L-j,a}$.
This operator commutes with the Hamiltonian at zero vector potential, $[M_x,H(0)]=0$.
It is noted that the ground state eigenvalues of $M_x, \tilde{M}_x$ depend on signs of $t,w$
in each phase because energies of bonding or anti-bonding states depends on the signs.
On the other hand, the index $\Delta p$ is independent of the signs.

The BBH model is a four-orbital spinless fermion model on the square lattice.
The hopping and hybridization are
\begin{align}
t_{j+\hat{x},j}^{\mu\nu}&=
\begin{pmatrix}
0 & 0 & 0 & 0\\
t & 0 & 0 & 0\\
0 & 0 & 0 & t\\
0 & 0 & 0 & 0\\
\end{pmatrix}, \quad
t_{j+\hat{y},j}^{\mu\nu}=
\begin{pmatrix}
0 & 0 & 0 & 0\\
0 & 0 & 0 & 0\\
0 & t & 0 & 0\\
t & 0 & 0 & 0\\
\end{pmatrix},\nonumber\\
w_{j}^{\mu\nu}&=
\begin{pmatrix}
0 & w & 0 & w\\
w & 0 & w & 0\\
0 & w & 0 & w\\
w & 0 & w & 0\\
\end{pmatrix}.
\end{align}
The $C_4$-rotation about the origin $(0,0)$ in absence of an external vector potential is
$C_4c_{j\mu}C_4^{-1}=R_{\mu\nu}c_{j'\nu}, j'=C_4j=(L-y_j,x_j)$,
\begin{align}
R_{\mu\nu}=
\begin{pmatrix}
0 & 1 & 0 & 0\\
0 & 0 & 1 & 0\\
0 & 0 & 0 & 1\\
1 & 0 & 0 & 0\\
\end{pmatrix}.
\end{align}
Although this model is well defined, it does not have a band gap at half-filling.
One need to introduce a flux $\theta_t\neq 0$ for each inter-site plaquette 
as a model parameter 
(which should be distinguished
from the external vector potential $A_{jk}$)
to create a stable band gap. We add such a flux in the gauge shown in Fig.~1 (c) and also introduce 
a flux $\theta_w$ for each intra-site plaquette in a $C_4$ symmetric way, $w\to we^{i\theta_w/4}$.
For simplicity, we consider $\theta_t=\theta_w=\theta$.
The quadrupole phase is extended for 
$0<\theta_{}\leq \pi$~\cite{Benalcazar2017Science,Benalcazar2017,Wheeler2019}.
The conventional $C_4$-rotation operator in absence of the external vector potential corresponding to
the flux $2\pi/L^2=O(L^{-2})$ in each plaquatte is replaced as
$C_4\to C_4\exp(i\theta_t\sum_jx_jy_jn_j)$.
This operator is the $C_4$-rotation symmetry operator of the reference Hamiltonian $H(0)$
in absence of the external vector potential $A_{jk}$
and we denote it simply as $C_4$.

\section{Properties of combined mirror and $C_4$-rotation operators}
\label{app:operator}

The square of $\tilde{M}_x$ for a one-dimensional system is 
\begin{align}
(\tilde{M}_x)^2
&=\exp\left(i\frac{2\pi}{L}\sum_{x_j=1}^{L-1}Ln_{L-j}\right) \nonumber\\
&\quad \times\exp\left(-i\frac{2\pi}{L}\sum_{x_j=1}^{L-1}(L-x_j)n_{L-j}\right)\cdot U_x \nonumber \\
&=1,
\end{align}
where $2\pi\sum_{j=1}^{L-1}n_j=0$ (mod $2\pi$) has been used for an integer filling $\rho$.
(Note that $n_j=\sum_{\mu}c_{j\mu}^{\dagger}c_{j\mu}-\rho$.)
Clearly, this holds true for any $L$ and also in higher dimensions.

For an odd $L$, there is a center site $x_0=0$ for the mirror operation under the open boundary condition
and it is convenient to introduce the coordinate $X_j=-(L-1)/2,\cdots,-1,0,1,\cdots,(L-1)/2$, where
$X_j=x_j$ for $0\leq x_j\leq (L-1)/2$
and $X_j=x_j-L$ for $(L+1)/2\leq x_j\leq L-1$.
In this coordinate, $X_j=-X_{L-j}$ under the mirror operation $M_x$.
Therefore, for $\tilde{M}_x'(s)=M_xU_x'(s)=M_x\exp(is2\pi/L\sum_{j=0}^{L-1}X_jn_j)
=M_x\exp(is2\pi/L\sum_{j=1}^{L-1}X_jn_j)$ with $0\leq s\leq 1$,
\begin{align}
(\tilde{M}_x'(s))^2
&=\exp\left(-is\frac{2\pi}{L}\sum_{j=1}^{L-1}X_{L-j}n_{L-j}\right)\cdot U_x'(s) \nonumber\\
&=1.
\end{align}
It is obvious that $\tilde{M}_x'(0)=M_x$ and $\tilde{M}_x'(1)=\tilde{M}_x$, which is the key in the proof of
the bulk-boundary correspondence in the main text.

Similarly, the square of $\tilde{C}_4$ is 
\begin{align}
(\tilde{C}_4)^2&=(C_4)^2C_4^{-1}U_{xy} C_4U_{xy} \nonumber\\
&=C_2\exp\left(i\frac{2\pi}{L}\sum_{x_j,y_j=1}^{L-1}x_jn_{x_j,y_j}\right) \nonumber \\
&\equiv C_2U_2\equiv \tilde{C}_2
\end{align}
The square of $\tilde{C}_2$ is evaluated similarly to that of $\tilde{M}_x$,
\begin{align}
(\tilde{C}_2)^2
&=\exp\left(i\frac{2\pi}{L}\sum_{x_j,y_j=1}^{L-1}Ln_{L-x_j,L-y_j}\right) \nonumber\\
&\quad \times\exp\left(-i\frac{2\pi}{L}\sum_{x_j,y_j=1}^{L-1}(L-x_j)n_{L-x_j,L-y_j}\right)\cdot U_2 \nonumber \\
&=1,
\end{align}
where $2\pi\sum_{x,y=1}^{L-1}n_j=0$ (mod $2\pi$) has been used for an integer filling $\rho$.
This gives $(\tilde{C}_4)^4=1$ for any $L$.

\begin{figure}[tbh]
\includegraphics[width=4.0cm]{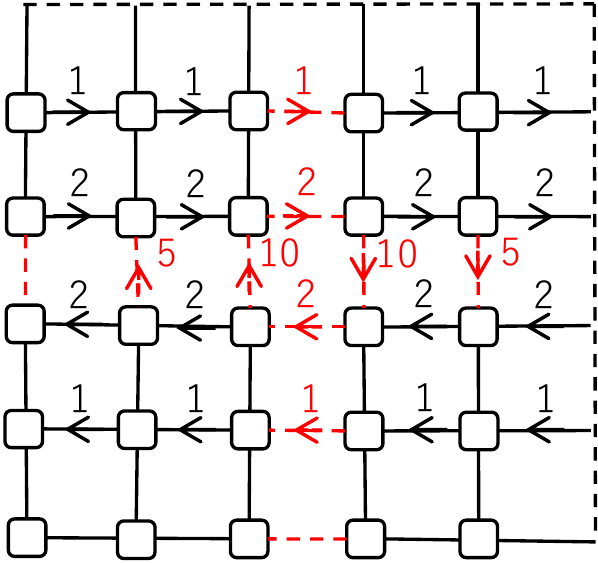}
\caption{The new gauge for $L=5$. The arrows with numbers (in unit of $2\pi/L^2$) represent $A_{jk}$.
The bonds with broken red lines have the hopping $t'$.
Note that, under the open boundary condition $t'=0$, 
this gauge admits a uniform flux $2\pi s/L^2$ when the parameter $0\leq s\leq 1$ is
introduced as $A_{jk}\to sA_{jk}$.
}
\label{fig:gauge_odd}
\end{figure}

When the linear system size $L$ is odd, there is a center site for the rotation operation under the open boundary
condition and it is convenient to introduce the coordinate $X_j,Y_j=-(L-1)/2,\cdots,0,\cdots, (L-1)/2$ as before.
They behave under the rotation as $C_4:(X_j,Y_j)\to (X_{C_4j},Y_{C_4j})=(-Y_j,X_j)$.
Correspondingly, we make a gauge transformation by ${\mathcal U}=\exp(i\sum_jf_jn_j)$ with 
$f_j=0$ for $Y_j\geq0$ and $f_j=(2\pi/L)X_j$ for $Y_j<0$ as mentioned in the main text.
Then, the combined symmetry operator becomes $\tilde{C}_4= C_4U_{xy}$ 
with $U_{xy}=\exp(i2\pi/L^2\sum_jX_jY_jn_j)$, where we have suppressed ``$f$" in $\tilde{C}_4^f$ and 
$U_{xy}^f$ for simplicity.
Note that a unifrom magnetic flux $2\pi s/L^2$ 
is realized under the open boundary condition $t'=0$, when the parameter $0\leq s\leq 1$ is introduced
as $A_{jk}\to sA_{jk}$ (Fig.~\ref{fig:gauge_odd}).
A straightforward calculation gives, 
for $\tilde{C}_4'(s)=C_4U_{xy}'(s)=C_4\exp(is2\pi/L^2\sum_jX_jY_jn_j)$ with $0\leq s\leq 1$,
\begin{align}
(\tilde{C}_4'(s))^4=1,
\end{align}
because of the rotation response of $(X_j,Y_j)$ mentioned above.
It is clear that $\tilde{C}_4'(0)=C_4$ and $\tilde{C}_4'(1)=\tilde{C}_4$ in the new gauge,
and they have the common eigenvalues for each of $s=0$ and $s=1$.

\section{Calculation of $\Delta q$ for BBH model}
\label{app:dq}

The ground state wavefunction for the topologically non-trivial case at $t<0, w=0$ is given
by Eq. (13) in the main text, where
the operator $\gamma_{jn}$ depends on the plaquette position $j$.
Thus we consider four disjoint regions of the system,
(i) $0\leq x_j,y_j\leq L-2$, (ii) $x_j=L-1, 0\leq y_j\leq L-2$, (iii) $0\leq x_j\leq L-2,y_j=L-1$, 
and (iv) $x_j=y_j=L-1$. 
When we write $\gamma_{j0}$ as $\gamma_{j0}^{\dagger}=1/2(u_{ja}c_{ja}^{\dagger}+u_{jb}c_{j+\hat{x},b}^{\dagger}
+u_{jc}c_{j+\hat{x}+\hat{y},c}^{\dagger}+u_{jd}c_{j+\hat{y},d}^{\dagger})$,
we have for the regions (i)$\sim$(iv),
\begin{align}
\begin{aligned}
u_{j\mu}^{\rm (i),(ii)}&=(\omega^{1/4},\omega^{-y_j},\omega^{-y_j-1/4},\omega^{1/2}), \\
u_{j\mu}^{\rm (iii),(iv)}&=(\omega^{-Lx_j+1/4},\omega^{-Lx_j-y_j},\omega^{3/4},\omega^{1/2}),
\end{aligned}
\end{align}
where $\omega=\exp(i2\pi/L^2+i\theta)$.
These are obtained by suitable gauge transformations of $u_{(0,0),\mu}$.
Similarly, for $\gamma_{j1}^{\dagger}=1/2(v_{ja}c_{ja}^{\dagger}+v_{jb}c_{j+\hat{x},b}^{\dagger}
+v_{jc}c_{j+\hat{x}+\hat{y},c}^{\dagger}+v_{jd}c_{j+\hat{y},d}^{\dagger})$, we have
\begin{align}
\begin{aligned}
v_{j\mu}^{\rm (i),(ii)}&=(\omega^{1/4},i\omega^{-y_j},-\omega^{-y_j-1/4},-i\omega^{1/2}), \\
v_{j\mu}^{\rm (iii),(iv)}&=(\omega^{-Lx_j+1/4},i\omega^{-Lx_j-y_j},-\omega^{3/4},-i\omega^{1/2}).
\end{aligned}
\end{align}
Then, a strightforward calculation gives
\begin{align}
\begin{aligned}
&{\rm (i)}\quad \tilde{C}_4\gamma_{jn}^{\dagger}\tilde{C}_4^{-1}
=\omega^{x_jy_j}\omega^{x_j}\omega^{1/4}e^{-i\pi n/2}\gamma_{j'n}^{\dagger}, \\
&{\rm (ii)}\quad \tilde{C}_4\gamma_{jn}^{\dagger}\tilde{C}_4^{-1}
=\omega^{x_jy_j}\omega^{x_j}\omega^{L(L-y_j-1)}\omega^{1/4}e^{-i\pi n/2}\gamma_{j'n}^{\dagger}, \\
&{\rm (iii)}\quad \tilde{C}_4\gamma_{jn}^{\dagger}\tilde{C}_4^{-1}
=\omega^{x_jy_j}\omega^{-x_jy_j}\omega^{1/4}e^{-i\pi n/2}\gamma_{j'n}^{\dagger} \\
&{\rm (iv)}\quad \tilde{C}_4\gamma_{jn}^{\dagger}\tilde{C}_4^{-1}
=\omega^{x_jy_j}\omega^{-x_jy_j}\omega^{1/4}e^{-i\pi n/2}\gamma_{j'n}^{\dagger},
\end{aligned}
\end{align}
where $j'=(L-y_j-1,x_j)$.
Therefore, the index $\Delta q$ is evaluated as
\begin{align}
2\pi \Delta q&=(2-\rho) \frac{2\pi}{L^2}\sum_{x,y=0}^{L-1}xy
+2\frac{2\pi}{L^2}\sum_{x=0}^{L-1}\sum_{y=0}^{L-2}x \nonumber\\
&\quad +2\frac{2\pi}{L^2}\sum_{y=0}^{L-2}L(L-y-1)
-2\frac{2\pi}{L^2}\sum_{x=0}^{L-1}x(L-1) \nonumber\\
&\quad +2\frac{2\pi}{L^2}\sum_{x,y=0}^{L-1}\frac{1}{4} \nonumber\\
&=2\pi\times \frac{1}{2} \quad (\mbox{mod } 2\pi).
\end{align}

\bibliography{ref}


\end{document}